\documentclass{pasj00}

\newcommand{\bm}[1]{\mbox{\boldmath$#1$}}

\begin{document}
\SetRunningHead{T. Hamana \& S. Miyazaki}{A note on artificial
deformation in object shapes due to the pixelization}
\Received{2007/07/13}
\Accepted{2008/07/19}

\title{A note on artificial
deformation in object shapes due to the pixelization}

\author{Takashi \textsc{Hamana}, 
Satoshi \textsc{Miyazaki}\\
{National Astronomical Observatory of Japan, Mitaka, 
Tokyo 181-8588, Japan}}

\maketitle

\begin{abstract}
We qualitatively examine properties of artificial deformation in shapes
of objects (galaxies and stars) induced by the pixelization effects 
(also called as the aliasing
effects) using toy mock simulation images.
Two causes of the effects have been recognized: One is a consequence 
of observing the continuous sky with discrete pixels which is called as 
the first pixelization.
And the other, called the second pixelization, is a consequence of 
resampling a pixelized image onto an another pixel grid whose
coordinates are not perfectly adjusted to the input grid.
We pay a special attention to the latter because it might be a potential
source of a systematic noise in a weak lensing analysis.
In particular, it is found that resampling with rotation induces
artificial ellipticities in object shapes having a
periodic concentric-circle-shaped pattern.
Our major findings are as follows.
(1) Root-mean-square (RMS) of artificial ellipticities in object shapes 
induced by the
first pixelization effect can be as large as RMS$\gtrsim 10^{-2}$ 
if a characteristic size of objects (e.g., the FWHM) is smaller than 
twice of the pixel size. 
While for larger objects, it quickly becomes very small 
(RMS$\lesssim 10^{-5}$).
(2) The amplitude of the shape deformation induced by the second
pixelization effect depends on the object size.
It also depends strongly on an interpolation scheme adopted to
carry out resampling and on the grid size of the output pixels. 
The RMS of ellipticities in object shapes induced by the second
pixelization effect can be suppressed to well below $10^{-2~}$ if one
adopts a proper interpolation scheme (implemented in popular 
image processing softwares).
We also discuss an impact of the pixelization effects on a weak
lensing analysis.
\end{abstract}

%
%
\section{Introduction}
A precise shape measurement is of fundamental importance in astronomical
research, not only because the morphology of celestial bodies provides
us with their physical information, but also because tiny
deformation in shapes of distant galaxies caused by gravitational
lensing effect allows
us to explore foreground mass distribution (see Fort \& Mellier 1994;
Mellier 1999; Bartelmann \& Schneider 2001 for reviews).  Analyses of
weak gravitational lensing effect (e.g., the cosmic shear correlation)
especially require very precise shape measurement, since its amplitude
is of a few percents level (Refregier 2003 and references
therein). Therefore, any artificial shape deformation which may arise
during an observation as well as data reduction must be properly
understood and must be controlled down to a sufficiently small level.

One known artificial shape deformation is caused by the pixelization of
images, known as {\it pixelization effects} (also called as the {\it 
aliasing effect}).  
As pointed out by Rhodes et al. (2007), there are two causes of
the pixelization effect (see Fig. 6 of Rhodes et al. 2007 for an
illustration): One is a consequence of observing the continuous sky with
discrete pixels, called as the {\it first pixelization}, which is an
unavoidable effect.  And the other, called as the {\it second
pixelization}, occurs when resampling a pixelized image onto an another
pixel grid whose coordinate is not perfectly adjusted to the input
grid. During resampling, one input pixel may be resampled onto
several output pixels, obviously resulting in deformation in object
shapes (see Figure \ref{fig:illust} for an illustration).  
The second pixelization occurs several stages of image processing
processes involving resampling, for example, the correction for a
geometric  distortion, mosaicking images from multiple CCD chips to
generate a combined image, and stacking multiple dithered images. 

In this paper, we are concerned with the second pixelization effect
taking two examples of resampling; rotation and correction for an
axially symmetric optical distortion. 
We especially pay an attention to rotation which must be involved in the
mosaic-stacking process  
of a mosaic CCD camera if multiple CCDs
are not installed in perfectly parallel with each other.  
In the case of the Subaru Prime
Focus Camera (Suprime-Cam), rotations of $0.025-0.17$ degrees are
necessary for generating a properly mosaicked image.
An important point to notice is that resampling with rotation induces
artificial ellipticities in object shapes having a
concentric-circle-shaped pattern (explained in
detail in the following sections and see Figures \ref{fig:illust} and
\ref{fig:demo} for demonstrations).  Therefore it give rise to
artificial shear correlations which can potentially act as a systematic
noise in the measurement of cosmic shear correlation functions.

We notice that the artificial shape deformation induced by the second
pixelization effect cannot be generally corrected by the anisotropic 
point spread function (PSF) correction (which is one of the important 
procedures in weak lensing analyses [Kaiser et al. 1995]) 
simply because they originate from different causes.
The shape deformation by the anisotropic PSF arises during an observation
with a real (thus non-perfect) instrument, thus it is an unavoidable
effect and one has to develop a reliable correction scheme (e.g., Heymans et
al. 2006; Massaey et al. 2007).
Whereas the second pixelization effect occurs during image processing,
and can be minimized by adopting an optimal resampling scheme. 
In fact, Rhodes et al (2007) developed such a resampling scheme 
for the HST ACS data in an empirical manner by searching for optimal 
parameters (the interpolation kernel and output pixel size) of the 
image processing software {\it MultiDrizzle}\footnote{see 
MultiDrizzle web page:\\ 
{\tt http://stsdas.stsci.edu/pydrizzle/multidrizzle/}}.

The purpose of this paper is two-fold: The first is to quantitatively
examine the effect of the second pixelization effect to understand its 
properties. The second is to explore
an optimum way to minimizing it.  To do these, we use simple image
simulations which is described in \S 2.  Then in \S 3, we give some
illustrative examples for a visual impression and for demonstrating the
origin of the concentric-circle-shaped pattern induced by resampling
with rotation.  Results are presented
in \S 4.  Finally, \S 5 is devoted to a summary and discussion.

%
%
\section{Simple image simulation}
\label{sec:simulation}

Since a very realistic mock simulation is not necessary for our purpose,
we use a toy image simulation described below.  
We adopt two-dimensional Gaussian as a shape of ``object''.
The full-width-half-maximums (FWHM denoted by $\theta_G$) of the Gaussian
object we consider are $\theta_G = 0.4$, 0.6, 0.8, 1.2 and 2.0 arcsec.
These object sizes are chosen because (i) the median
seeing size (FWHM) of the Subaru telescope is about 0.6 arcsec and the best
seeing is $\sim 0.35$ arcsec (Miyazaki et al. 2002), and (ii) most of
objects used for weak lensing analyses are galaxies (and reference
stars) with the FWHMs being smaller than 2 arcsec 
(Hamana et al. 2003).  

We take the same pixel size as the Suprime-Cam, namely 
$\l_{\rm pixel}=0.2$ arcsec,  as our primary
science target is the weak lensing, especially using the Suprime-Cam or
similar instruments.
We create mock CCD images having $N_x \times N_y$ pixels on which
Gaussian objects (having the equivalent FWHM and intensity) are located
on a regular interval of $3 \pi$ arcsec.  
Note that the separation between objects are more than 10 times of the 
$\sigma$ of the Gaussian ($\theta_G \simeq 2.35 \times \sigma$), thus 
overlapping of isophotes of neighbour objects does not make any problem
in the shape measurement.
Note that results in this paper can be
applied to any camera that uses a pixel array imaging device
by properly translating the scaling ratio
between $\theta_G$ and $l_{\rm pixel}$.

\begin{figure}
\includegraphics[height=85mm]{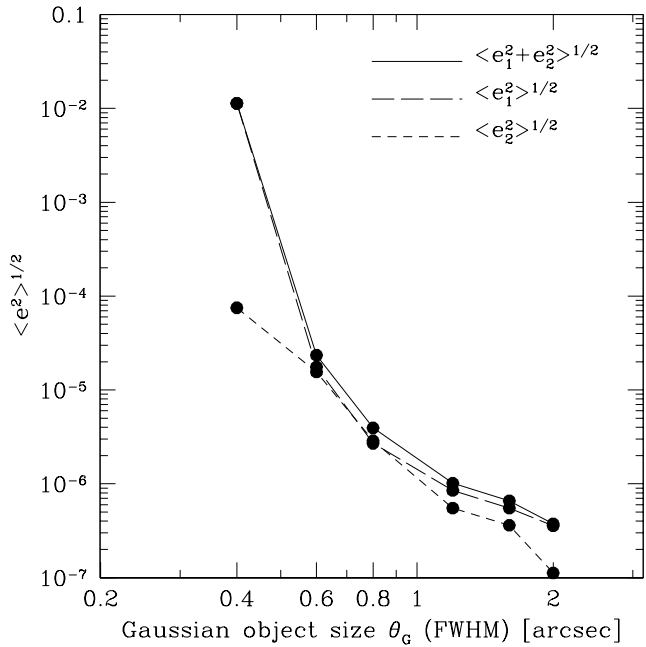}
\caption{The RMSs of the ellipticities in object shapes caused by the first 
pixelization effect is plotted as a function of the FWHM of the Gaussian
objects. 
Note that the RMS ellipticities of stars before the PSF correction
  is typically a few percents, and the RMS of intrinsic galaxy
  ellipticities is about 40 percents (e.g., Hamana et al. 2003).
\label{fig:1st}}
\end{figure}

Following the, so-called, KSB formalism (Kaiser, Squires \& Broadhurst
1995), we quantify the image shapes by the ellipticity parameter defined
by 
\begin{equation}
\label{eq:e}
\bm{e}=\left({{I_{11} - I_{22}}\over {I_{11}+I_{22}}},
{{I_{12}}\over {I_{11}+I_{22}}} \right),
\end{equation}
\begin{equation}
\label{eq:I}
I_{ij}=\int d^2\theta ~ W_G(\theta) \theta_i \theta_j f(\bm{\theta}),
\end{equation}
where $W_G(\theta)$ is the Gaussian window function.
Notice that the $e_1$ ($e_2$) component represents the elongation in
directions parallel (45 degrees rotated) to the coordinate system.
The object detection and shape measurement are done with {\it
  hfindpeaks} and {\it getshapes} of IMCAT software suite developed 
by Nick Kaiser, respectively.

\begin{figure*}
\begin{center}
\includegraphics[height=85mm]{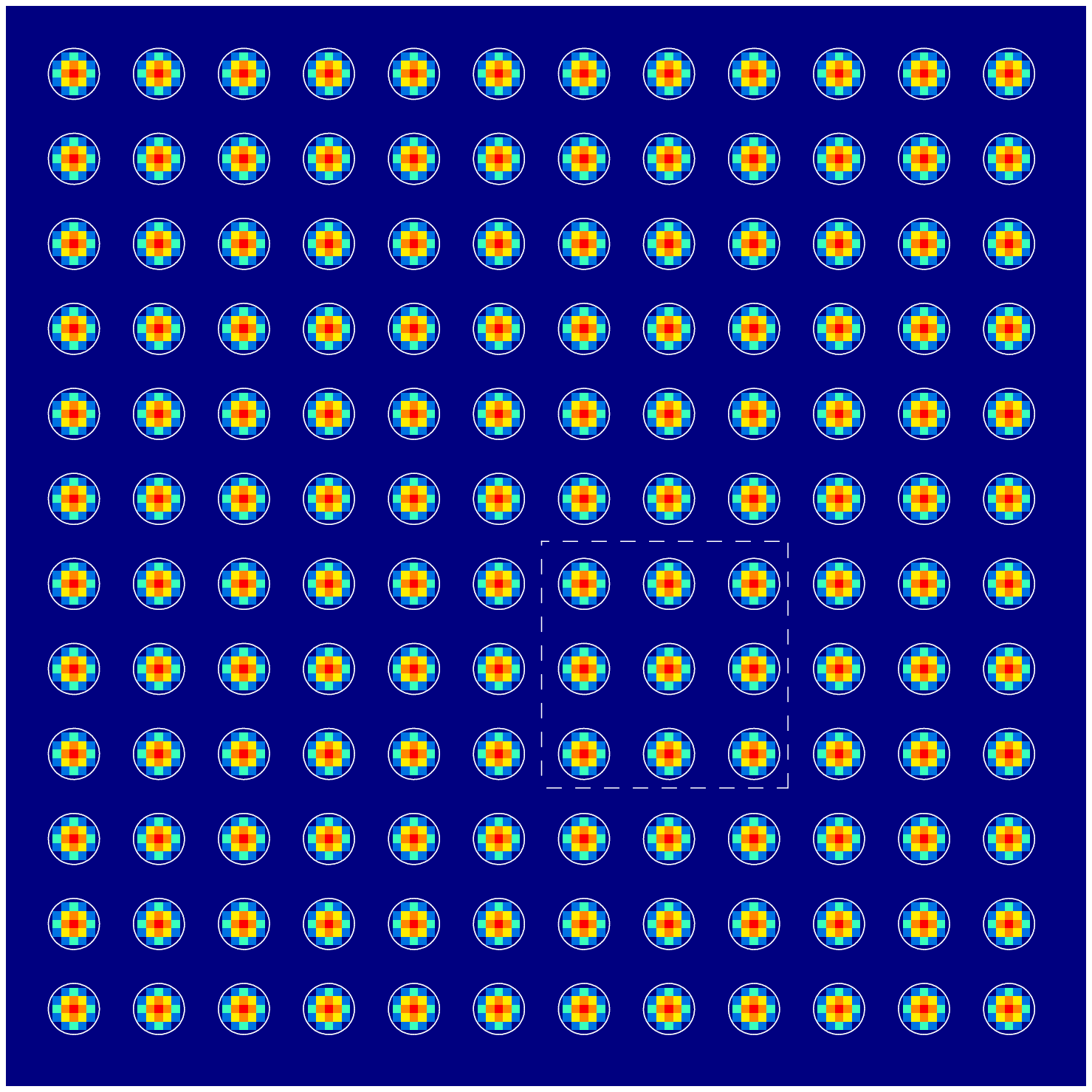}
\hspace{2mm}
\includegraphics[height=85mm]{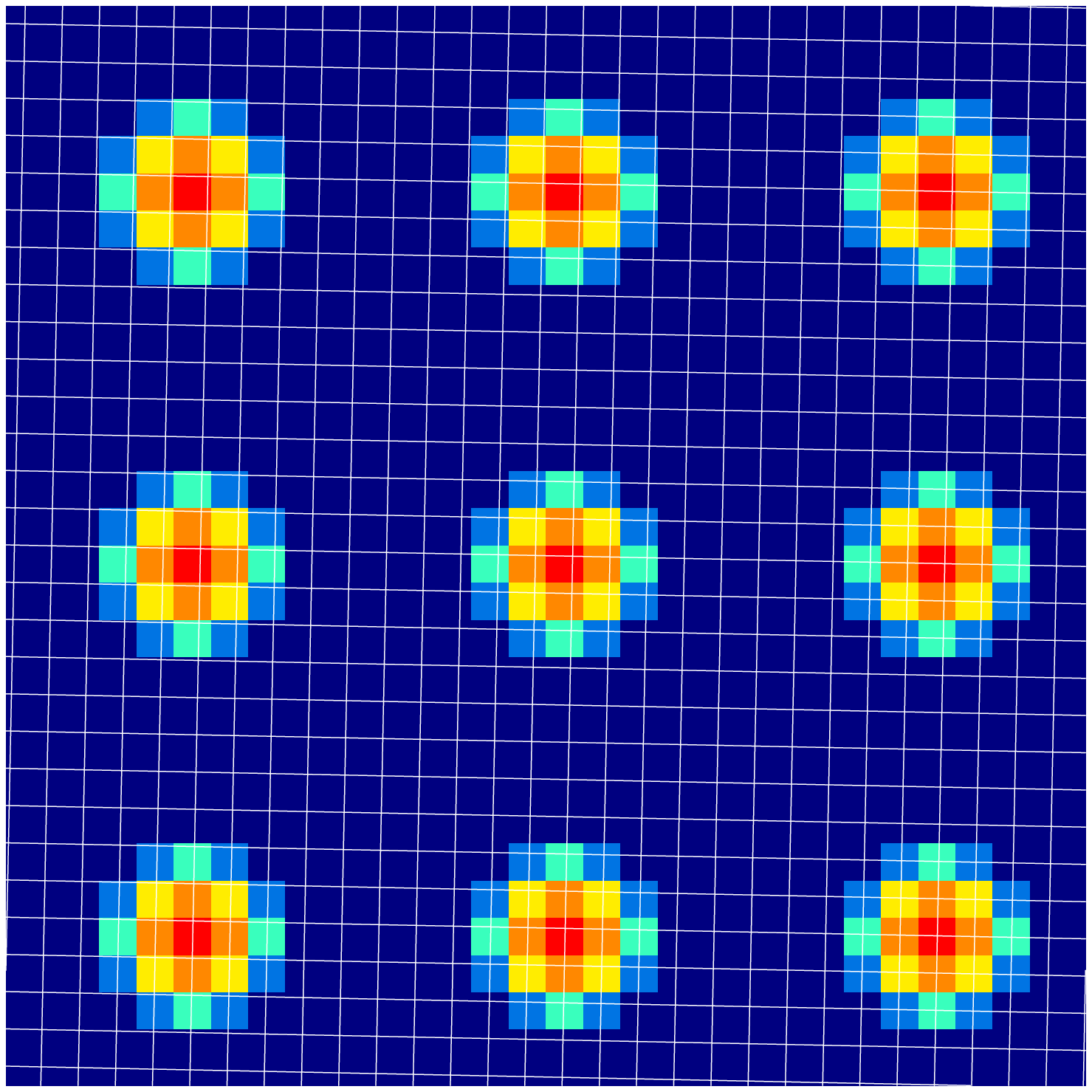}\\
\vspace{2mm}
\includegraphics[height=85mm]{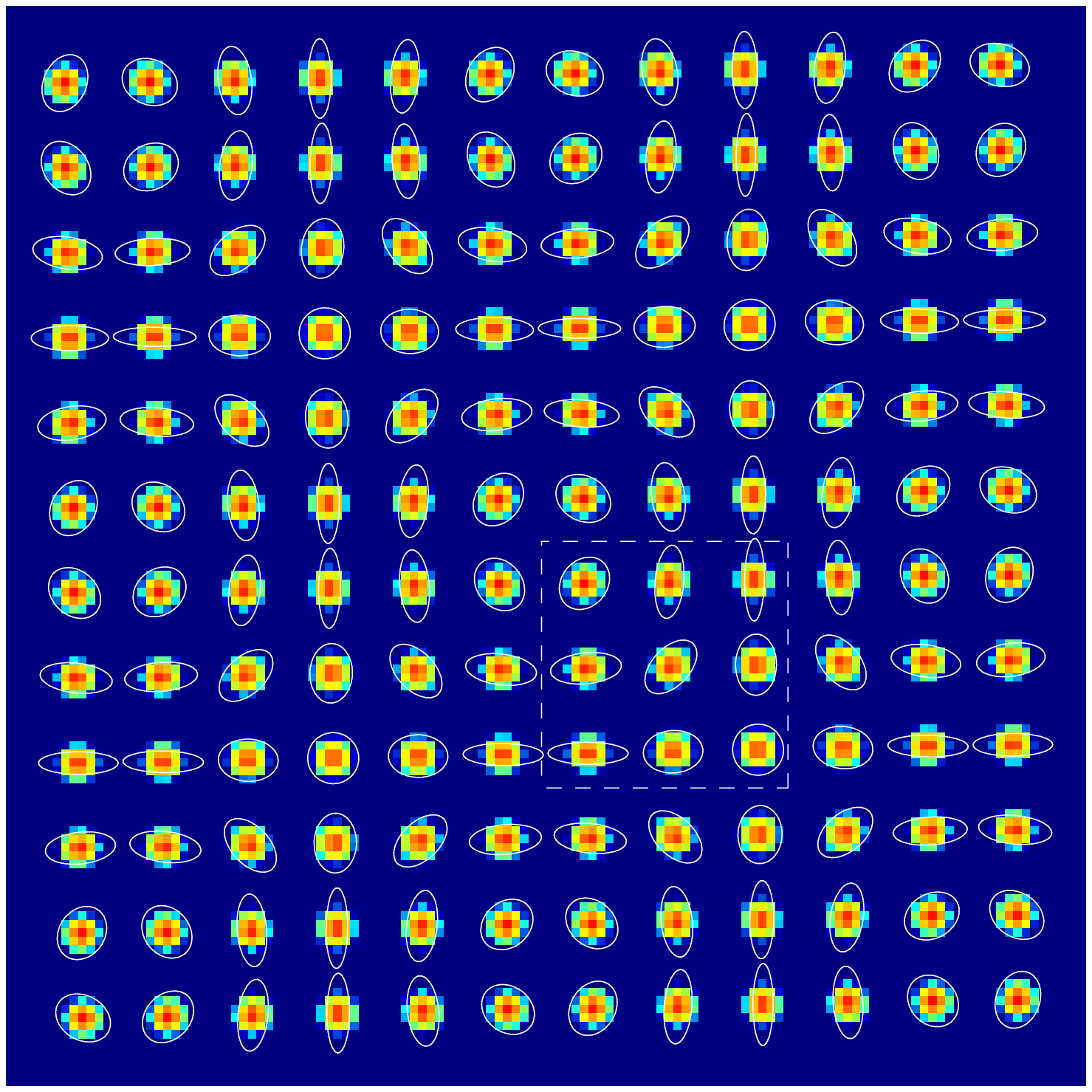}
\hspace{2mm}
\includegraphics[height=85mm]{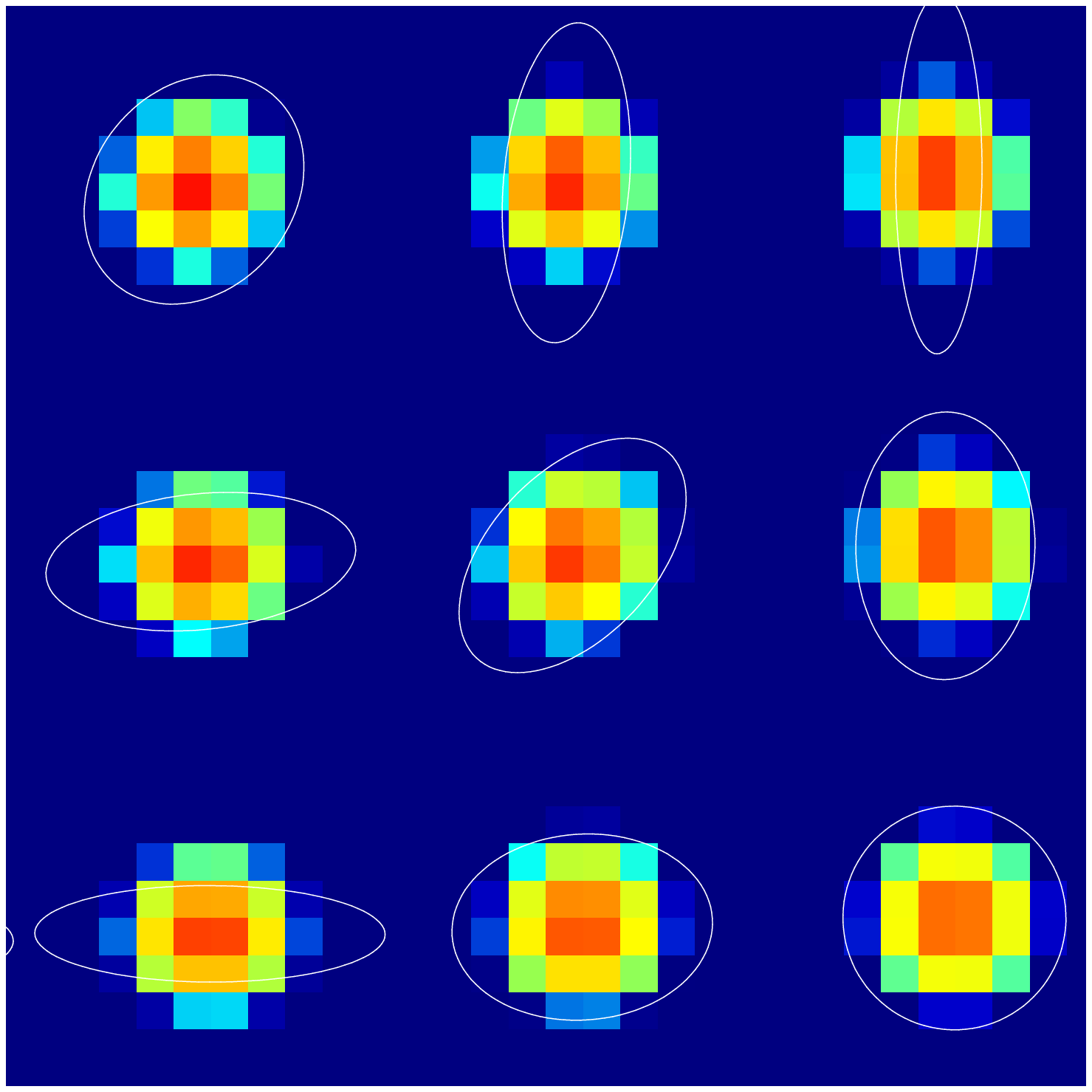}
\\
\vspace{2mm}
\includegraphics[height=85mm,angle=-90]{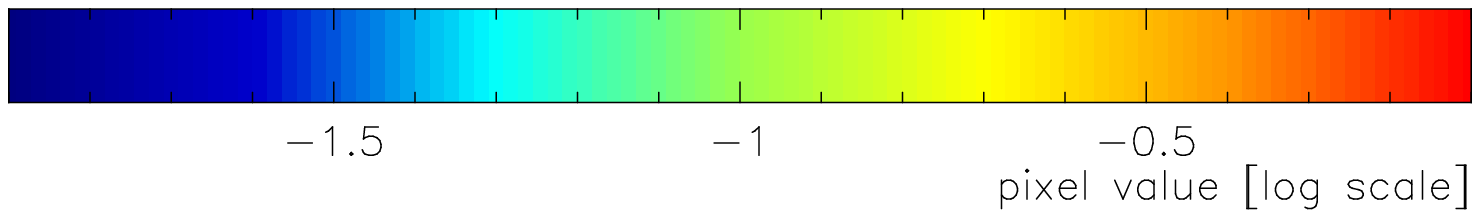}
\end{center}
\caption{An illustrative example of the object shape deformations caused 
by the second pixelization. 
Top-left panel shows the simulation data of Gaussian objects with 
$\theta_G/ l_{\rm pixel} = 2$ located on a regular interval of 10
pixels.
Note that the objects are placed just at the center of pixels to
minimize the shape deformation caused by the first pixelization effect.
The top-right panel shows the zoom-in on the section enclosed by the
dashed line in the top-left panel. In this plot, a new grid (rotated 
by 1.15 degrees relative to the original grid) on which the image is
resampled is over-plotted.
The bottom two panels show the image after resampling onto the new grid.
The resampling is done with the 1st order bilinear polynomial
interpolation scheme.
The over-plotted ellipses show the ellipticity of the objects.
Note that the ellipticities are enlarged 10 times for clarify.
\label{fig:illust}}
\end{figure*}

Before investigating the second pixelization effect, 
here we examine the {\it first pixelization} effect. 
The Gaussian has, of course, no ellipticity, but its pixelized image may 
have a finite ellipticity if the center of an object does not fall onto
special positions such like the center of a pixel or an intersection of
grid.
We compute the root-mean-square (RMS) of the ellipticities among 
Gaussian objects in the simulation data of $2048\times 2048$ pixels.
The RMS is defined by
\begin{equation}
\label{eq:RMSe}
\langle e^2 \rangle^{1 \over 2} = 
\left[ \sum_{i=1}^{N_{obj}} (e_{i,1}^2+e_{i,2}^2)/N_{obj} \right]^{1\over 2}.
\end{equation}
Results are plotted in Figure \ref{fig:1st}.  As expected, the RMS
decreases with the object size.  It quickly becomes large for smaller
objects of $\theta_G <0.6$ arcsec and is $\sim 0.01$ for the case of
$\theta_G = 0.4$ arcsec ($\theta_G/l_{\rm pixel}=2$).
This should be compared with the RMS ellipticities of stars (before the 
anisotropic PSF correction) which is typically a few percents. 
Thus if the seeing FWHM is less than twice of the pixel size, 
the first pixelization effect can be one of major sources of artificial shape
deformation in small objects.
An important finding here is that for objects
with FWHM larger than three times the pixel size, the first pixelization
effect is very small, but for smaller objects, the first pixelization
effect can be a non-negligible source of the artificial ellipticities.
Another point to be noticed here is that the pixelization effect does
not necessarily generate the RMSs of $e_1$ and that of $e_2$ equally,
because the pixels are square shaped and so the pixelization effect is,
in general, not axially symmetric but has some special directions.
This is the reason why the RMSs of two components are, in general, not
equal as shown in Figure \ref{fig:1st}.

%
%
\section{Visual impressions}

\begin{figure}
\begin{center}
\includegraphics[width=85mm]{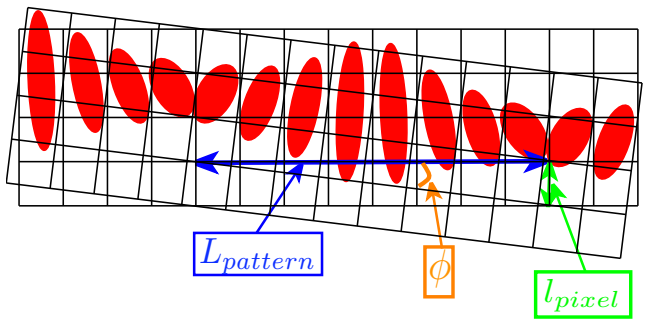}
\end{center}
\caption{A sketch explaining the relation between 
the scale of the pattern ($L_{pattern}$), the pixel size 
($l_{pixel}$) and the rotation angle ($\phi$).
Two grids show the input and output grids which cross at an
angle of $\phi$.
The ellipses show the ellipticites (which are 
enlarged arbitrarily for clarify) of originally circular objects
resampled onto the output grid.
Notice that the same deformation pattern appears at the interval of 
$L_{pattern}$.
\label{fig:sketch}}
\end{figure}

Before moving on to a thorough examination of the 
second pixelization effect, it would be helpful to present some
illustrative examples of the second pixelization effect.
Notice that in this section, for the illustrative purpose, we use
simulation data which are different from ones used in \S 4 in
the separation between objects.
Figure \ref{fig:illust} shows a demonstrative example for the origin of 
the periodic pattern caused by a resampling with a rotation.
Here, we create simulation  data of Gaussian
objects with $\theta_G/ l_{\rm pixel} = 2$ located on a regular interval
of 10 pixels.  
Note that the objects are placed exactly 
at the center of pixels to minimize the ellipticity induced by
the first pixelization effect.  The top-left panel shows the simulation
image where ellipticities of object shapes are over-plotted by ellipses
(circles in this case).  The top-right panel shows the zoom-in on the section
enclosed by the dashed line in the top-left panel. In this plot, a new
grid (rotated 1.15 degrees relative to the original grid) on which the
image is resampled is over-plotted.  
Note that the rotation angle of 1.15 degrees is much larger than a usual 
rotation angle involved in the mosaicking of multiple CCDs, 
but is chosen for the demonstrative purpose.
The bottom two panels show the
images after resampling onto the new pixels.  It is evident from these
plots that a periodic pattern of artificial ellipticities in object
shapes are induced by the rotation.  
The characteristic scale of the pattern is written
in terms of the pixel size and the rotation angle, $\phi$, as
$L_{\rm pattern} = l_{\rm pixel} / \tan\phi$ 
($\sim 50 \times l_{\rm pixel}$ for $\phi=1.15$ degree).
The reason of this is as follows (see Figure
\ref{fig:sketch} for an illustration):
The deformation is induced by the difference in the grid positions
between the input and output grids.
Thus if the difference in the grid positions is same at separate positions, 
the same deformation is induced at those positions.
In $x$- and $y$-direction, the same difference in the grid positions
occurs in the interval of $L_{\rm pattern}$, because it is the length 
that the output grid diagonally crosses (with the angle of $\phi$) 
the input grid by one pixel length.
Thus $L_{\rm pattern}$ is the separation between positions (in $x$-
and $y$-direction) where the same deformation is induced.

\begin{figure*}
\begin{center}
\includegraphics[height=175mm,angle=-90]{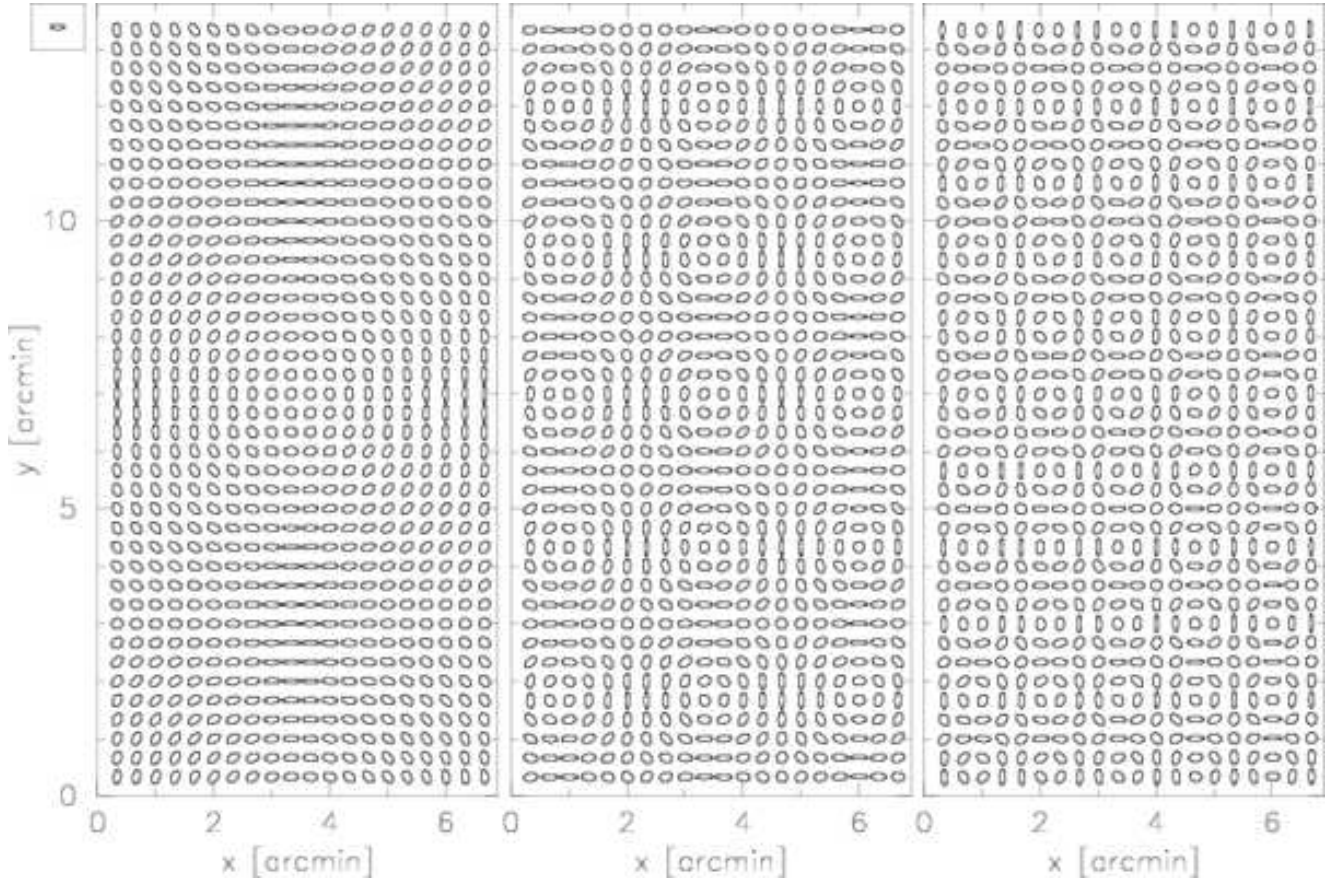}
\end{center}
\caption{Ellipticity maps showing the periodic patterns of object
shape deformations arising on realistic data.
Simulation data having the same dimensions as
CCDs of Suprime-Cam ($2048\times 4096$ pixels with $l_{\rm 
pixel}=0.2$ arcsec) on which Gaussian images of $\theta_G =0.6$ arcsec are 
placed on a regular interval of 20 arcsec, are rotated by 0.025, 0.075
and 0.15degrees (from left to right).
Note that the ellipticities are enlarged 20 times for clarify.
 For comparison, an ellipse with $|e|=2\%$ (enlarged 20 times) is 
displayed in the small panel at the top-left corner.
\label{fig:demo}}
\end{figure*}

Next, in order 
to demonstrate the periodic patterns of ellipticities in object shapes
appearing in realistic data, we create simulation data having the same
dimensions 
as CCDs of Suprime-Cam (namely, $2048\times 4096$ pixels with $l_{\rm
pixel}=0.2$ arcsec) on which Gaussian images of $\theta_G =0.6$ arcsec are
placed on a regular interval of 20 arcsec.  The data are rotated by
0.025, 0.075 or 0.15 degrees and are resampled onto new pixels by
adopting the 
1st order polynomial interpolation scheme (see \S 4 for details).
These rotation angles are chosen because the actual rotation  
involved in mosaicking of the Suprime-Cam's CCDs ranges from 0.025 to 0.17
degrees.
Ellipticity maps of resampled images are shown in Figure \ref{fig:demo},
where the characteristic periodic patterns are clearly observed.  The
scales of the pattern are $L_{\rm pattern} = 0.2'' / \tan\phi \sim$ 7.6,
2.5 and 1.3 arcmin for $\phi=$ 0.025, 0.075 and 0.15 degrees,
respectively.  This explains, at least qualitatively, the origin of the
concentric-circle-shaped pattern observed in the real data displayed in
Figure 2 of Miyazaki et al. (2007).  
As evidently shown in Figure
\ref{fig:demo}, the artificial ellipticities induced by the image
rotation mostly lead to the E-mode shear.  It is thus very important to
note that in the presence of such systematic ellipticities, a smallness
of the B-mode shear does not guarantee a successful 
correction of this systematic noise, and it may be difficult to
distinguish this
from signals arising from gravitational lensing.  
Thus it is necessary
to develop a resampling procedure which suppresses the systematic to 
a sufficiently small level.
This is exactly a purpose of this paper, and we explore the way to
minimising the systematic in an empirical
manner in the next section.
Notice that the actual mosaick-stacking involves
the rotation, displacement and enlargement/reduction of images, also a
high order warping is operated to remove the optical distortion (e.g.,
see Miyazaki et al. 2007). Thus actual data may have more complex
ellipticity pattern than ones found in the simple simulation in this
section.
In the next section, we shall qualitatively examine the second
pixelization effect taking two realistic examples of resampling; namely 
rotation operated in mosaicking and correction for the optical
distortion in the case of Suprime-Cam.

%
%
\section{Results}

The magnitude of the second pixelization effect depends on the
interpolation scheme used to resample an image.
We examine the following interpolation schemes which are
implemented in some popular image processing softwares: 
(i) the polynomial interpolation of 1st, 3rd and 5th order, for which
we utilize {\it transformimage} of IMCAT.
(ii) the sinc kernel [${\rm sinc}(x) = \sin(\pi x)/\pi x$] (truncated at
31 by 31 pixels), for which we utilize {\it rotate} of
IRAF\footnote{see IRAF web page {\tt http://iraf.noao.edu/}}.
(iii) Lanczos kernel 
[${\rm sinc}(x) {\rm sinc}(x /a)$] of $a=2$, 3 and 4 (called Lanczos2,
Lanczos3 and Lanczos4, respectively; implemented e.g.,
{\it Swarp} developed by Emmanuel Bertin), for which we utilize a
resampling program developed by ourself.
Also we examine the performance of adopting a finer
grid for output pixels, which we call the {\it grid refinement}.
Actually, it has been recognized that the grid refinement can
reduce object shape deformation by the pixelization effects (Rhodes et
al. 2007; Miyazaki et al. 2007) at the cost of the computational
overheads. 
The {\it grid refinement} was tested in combination with the 1st and 3rd
order polynomial interpolation schemes for which we utilize  {\it
  transformimage} of IMCAT. 

\subsection{Rotation}

The $2048 \times 2048$ pixel simulation data described in \S 2 are
rotated by 0.16 degrees and
are resampled onto a new grid by applying one of the interpolation schemes
mentioned above.  
The rotation angle of 0.16 degrees is chosen so that it is within
the range of Suprime-Cam's actual rotation angles in the 
mosaic-stacking procedure ($0.025-0.17$ degrees).
Note that the RMS of the ellipticities after resampling does not
depend on the rotation angle, though the size of
the concentric-circle-shaped pattern does.

\begin{figure*}
\begin{center}
\includegraphics[height=85mm]{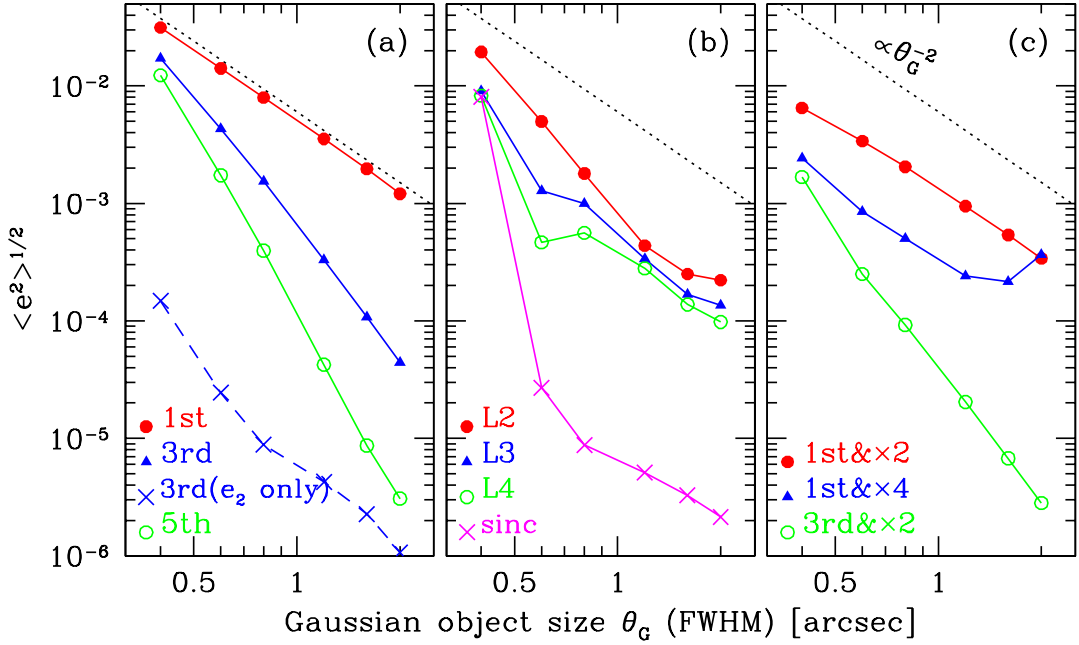}
\end{center}
\caption{RMSs of ellipticitis in object shapes as a function of the
object size are shown for comparison among various interpolation
schemes. 
{\it Left panel} (a):
the 1st, 3rd and 5th order polynomial for the filled circles, 
filled triangles and open circles respectively. The crosses show the
contribution from $e_2$ component ($\langle {e_2}^2 \rangle^{1/2}$) of 
the 3rd order polynomial, which demonstrates that the second
pixelization effect mostly induces $e_1$ component.
{\it Middle panel} (b):
the Lanczos2, Lanczos3, Lanczos4 and sinc resampling schemes for the
filled circles, filled triangles, open circles and crosses,
respectively. 
{\it Right panel} (c):
The 1st order polynomial with twice and 4 times finer grid for 
the filled circles and filled triangles respectively, and the 3rd
order polynomial with twice finer grid for the open triangles.
\label{fig:thetag}}
\end{figure*}

Let us first look into the dependence of the second pixelization effect
on the object size for various interpolation schemes.
Figure \ref{fig:thetag} compares the RMSs of
ellipticities in object shapes as a function of the object size.
Left panel of Figure \ref{fig:thetag} compares the three polynomial 
interpolation schemes,  revealing that
the higher the order of polynomials is, the better performance one
obtains.
It is also found that the higher the order
of  polynomials is, the steeper the slope becomes.
To be specific, the RMSs depend on the object size roughly,
$\langle e^2 \rangle^{1/2} \propto \theta_G^{-2}$ for the linear
polynomial, and $\langle e^2 \rangle^{1/2} \propto \theta_G^{-4}$ for
the 3rd order, and further steeper slope for the 5th order.
The crosses in the same plot show the RMS of $e_2$ component only for
the case of the 3rd order polynomial, from which it is found that the
$e_2$ component is much smaller than the total RMS.
In fact, we found that the second pixelization effect preferentially
induces $e_1$ component, irrespective of the interpolation schemes.
This is due to the fact that the pixels are square shaped and so the
pixelization effect has some special directions. 

The middle panel of Figure \ref{fig:thetag} shows the results for 
the sinc and Lanczos kernels. 
It is found that the Lanczos2 works as well as the 3rd order
polynomial does.
The Lanczos3 and Lanczos4 are better than 3rd and 5th order
polynomials for small objects ($\theta_G<0.6$ arcsec) but for larger
objects ($\theta_G>1$ arcsec) they work only a little better than
Lanczos2 does. 
The sinc kernel shows the best performance among the interpolation
schemes (without the grid-refinement) we consider in this paper.
We note that the sinc kernel is computationally expensive as it extends
to very large area (e.g., 31 by 31 pixels for the default setting of {\it
  IRAF}).

\begin{figure}
\begin{center}
\includegraphics[height=85mm]{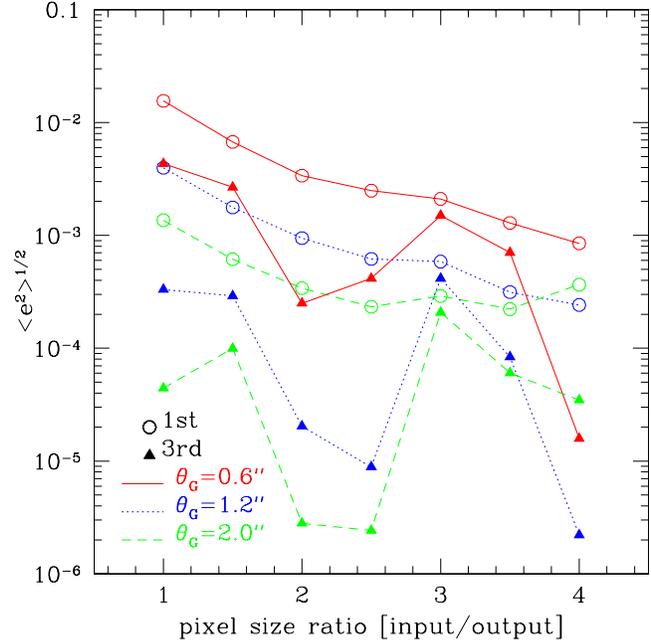}
\end{center}
\caption{RMSs of ellipticitis in object shapes as a function of the ratio
between the input and output pixel size.
Open circles are for the linear polynomial interpolation scheme, while
filled triangles are for the 3rd order polynomial.
Different line styles are different object size: $\theta_G=0.6$, 1.2 and
2.0 for the solid, dotted and dashed line, respectively.
\label{fig:sampling}}
\end{figure}

\begin{figure*}
\begin{center}
\includegraphics[height=89mm,angle=-90]{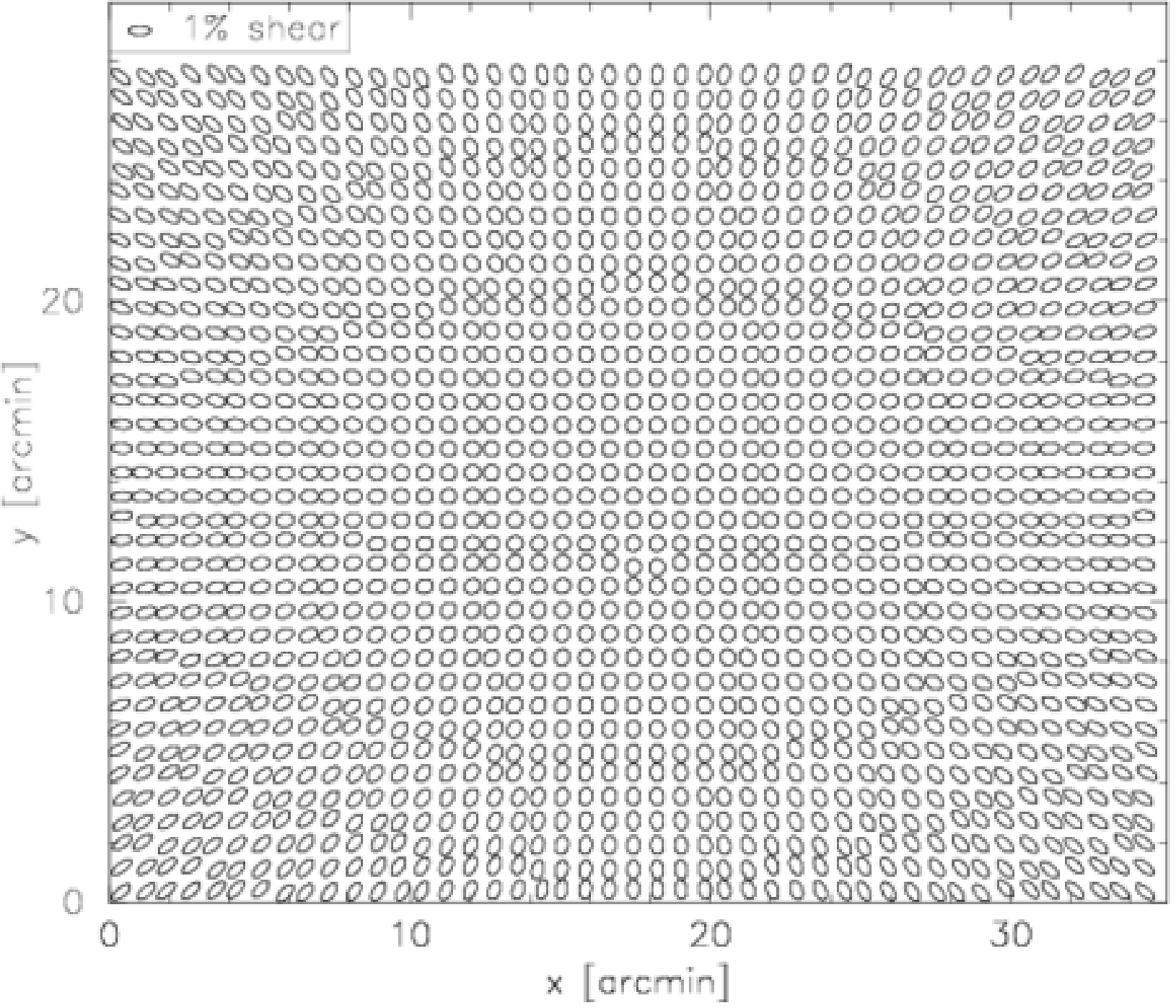}
\includegraphics[height=89mm,angle=-90]{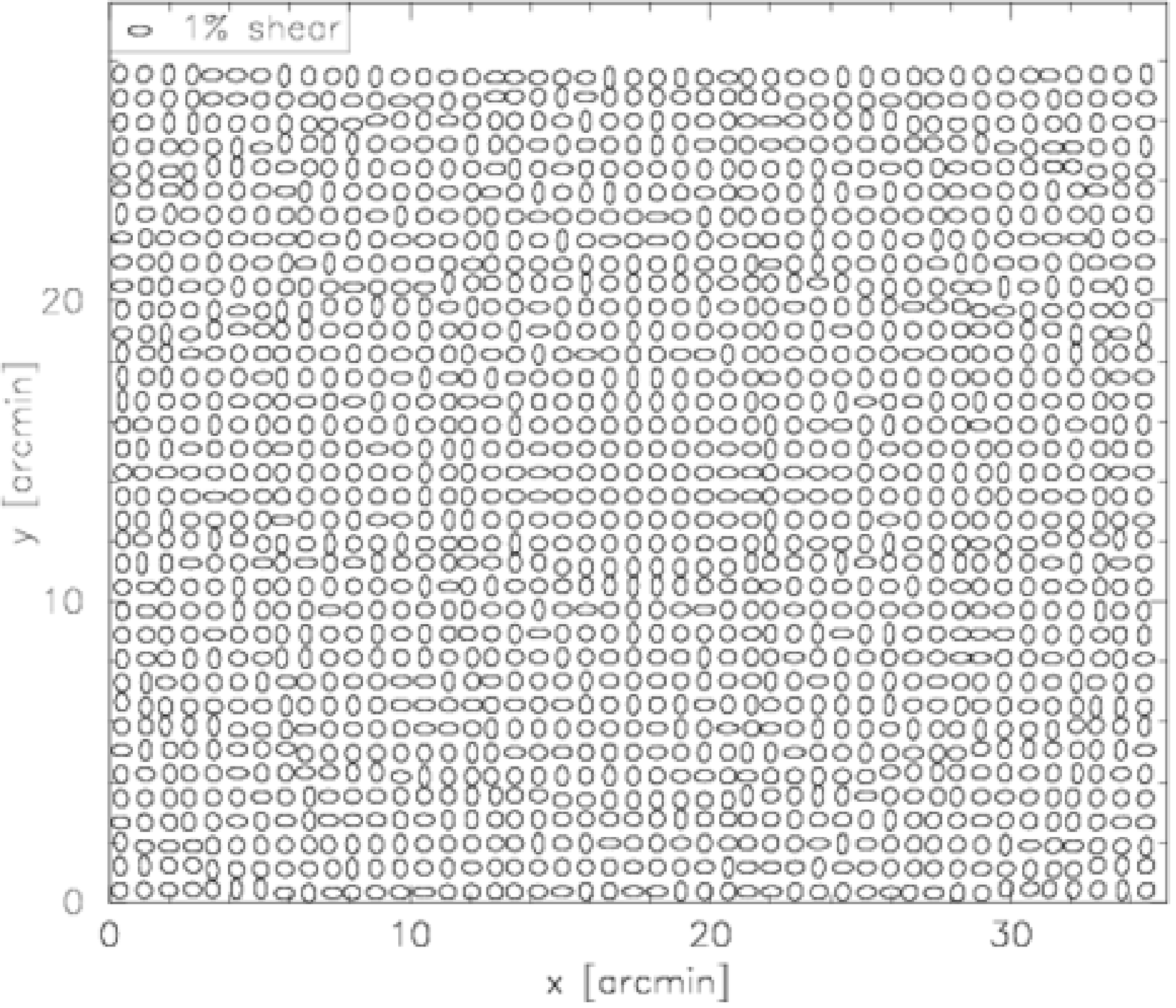}
\vspace{2mm}
\end{center}
\caption{Ellipticity maps before (left) and after (right) the distortion
  correction.
The left panel shows the mock simulation of the ellipticity map due to
  the optical distortion of the Suprima-Cam. 
The Gaussian objects with FWHM of 0.6 arcsec with the optical distortion 
are distributed on a mock Suprime-Cam pixels.
The ellipses show the ellipticity of distorted Gaussian objects sparsely
  sampled from the mock simulation data (see \S \ref{sec:simulation}).
The right panel shows the ellipticity map of the distortion corrected data.
For comparison, an ellipse with $|e|=1\%$ is 
displayed in the top-left corner.
\label{fig:dist-demo}}
\end{figure*}

Right panel of Figure \ref{fig:thetag} shows that the grid
refinement nicely suppress the second pixelization effect.
This is also observed in Figure \ref{fig:sampling} where the improvement 
gained by the grid refinement is plotted as a function of the ratio
between input and output pixel size.
If combined with the linear polynomial interpolation scheme, 
taking twice finer output grid reduces the RMSs by about one 
third with keeping the 
slope of $\langle e^2 \rangle^{1/2} \propto \theta_G^{-2}$ mostly unchanged.
The use of 4 times finer grid reduces the RMSs by about one order of
magnitude for objects with 
$\theta_G < 1.2$ arcsec and by lesser extent for larger objects. 
If combined with the 3rd order polynomial, improvements gained by
the grid refinement behave irregularly as observed in Figure
\ref{fig:sampling}.
Interestingly, in the case of an input/output pixel ratio of 3,
adopting the 3rd order polynomial makes only a slight improvement
over the 1st order case.
An important message of this is that certain combinations may not give 
good improvement for the computational overhead, and thus care must be
paid when one combines the grid refinement with a higher order
interpolation scheme.
Our experiment suggests that reasonably good improvement is stably
obtained when one adopts the twice finer grid with the 3rd order
polynomial interpolation.

%
%
\subsection{Optical distortion}
\label{sec:distrotion}

Next we examine the second pixelization effect induced during the
correction for the optical distortion.
To do so, we take the case for the Suprime-Cam.
The optical distortion of the Suprime-Cam is axially symmetric
with respect to the optical axis and is well approximated by the forth
order polynomial function of the distance from the optical axis (see
eq. [8] of Miyazaki et al. 2002). 
As is shown in Fig 21 of Miyazaki et al. (2002), the distortion rapidly
increases with the distance from the optical axis. 

\begin{figure}
\begin{center}
\includegraphics[height=85mm]{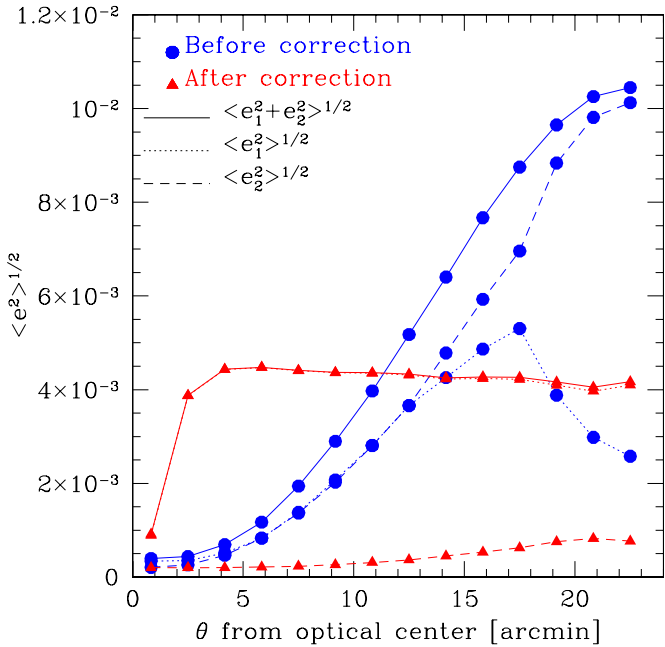}
\end{center}
\caption{RMSs of ellipticitis in object shapes as a function of the
distance from the optical axis.
This is the result of the mock Gaussian simulation with the FWHM of 0.6
arcsec.
Filled circles and filled triangles are for before and after distortion
correction. The dotted and dashed lines are for $e_1$ and $e_2$
components, respectively, whereas the solid lines show the sum of
them.
\label{fig:distrotion}}
\end{figure}

Adopting the forth order polynomial model given in Miyazaki et
al. (2002; their eq. [8]), we generate a mock Suprime-Cam images of
$10456\times 8282$ pixels, on
which Gaussian objects with $\theta_G=0.6$ arcsec with the optical
distortion artificially operated, are distributed in the manner
described in \S \ref{sec:simulation}.
In the left panel of Figure \ref{fig:dist-demo}, the ellipticity map of
the distorted Gaussian objects is shown. As is shown there, the
Suprime-Cam's optical distortion induces a radial elongation in the
object shapes because the distortion becomes larger as the distance
from the optical axis increases.
The RMS of the ellipticities as a function of the distance from the
optical axis is shown in Figure \ref{fig:distrotion}.
At the central region where the distortion is smallest, the RMS
ellipticity is as small as one induced by the first pixelization effect
as expected. Whereas at the largest distance it becomes one percent.
Note that in this case the first pixelization effect induces RMSs of
$e_1$ and  $e_2$ almost equally, and the turnover in $e_1$ component
seen at  $\theta \sim 17$ arcmin is an artifact due to the anisotropic
sampling (objects in the largest distance are located only at the four
corners, which preferentially have $e_2$ component as observed in Figure 
\ref{fig:distrotion}).

We correct for the optical distortion by resampling pixels with the 
3rd order polynomial interpolation. 
The ellipticity map after the correction is shown in the right panel of
Figure \ref{fig:dist-demo} from which one may visually realize that the
radial deformation induced by the optical distortion is corrected
successfully but artificial deformations due to the second pixlization
effect appear. 
The important point recognized in the plot is that the ellipses are
preferentially oriented to the directions parallel to the grids. 
This is quantitatively demonstrated in Figure \ref{fig:distrotion} in 
which one may find that the RMS of the ellipticities after the
correction almost solely comes from $e_1$ component.
Note that the RMS only weakly depends on the the distance from the
optical axis except for the most central region where the optical
distortion is negligible and thus the correction is as well. 
The values of the RMS (
$\langle e^2 \rangle^{1/2}\sim \langle {e_1}^2
\rangle^{1/2}\sim  4\times 10^{-3}$
and $\langle {e_2}^2 \rangle^{1/2}\sim 5\times 10^{-4}$) 
are equivalent to that
found in the case of the rotation (Figure \ref{fig:thetag}).
These similarities suggest that the RMS ellipticities induced by the second
pixelization effect does not depend on resampling parameters 
(e.g., displacements, rotation and anisotropic transformation) but
solely depends on the ratio between the object size and the pixel size.  
Although this could be a
specific feature of the Gaussian, it might be generally said that the
magnitude of the second pixelization effect most strongly depends on the
object size.

Notice that in a closer look at the left panel of Figure
\ref{fig:dist-demo} one may observe partially mirror symmetric patterns
with respect to the $x$- and $y$-axis passing through the field center.
The reason of this is that although the optical distortion and thus the
correction for it are axially symmetric, resampling onto square
pixels does not induce an axially symmetric pattern but results in the
partially mirror symmetric pattern. 
Except for it, we do not observe any obvious characteristic pattern.

%
%
\section{Summary and Discussions}
\label{sec:summary}

We have qualitatively examined ellipticities in object shapes induced by
the pixelization effects paying a special attention to the periodic
concentric-circle-shaped pattern induced by resampling of pixels 
with rotation.
Our major findings are summarized as follows.
\begin{itemize}
\item Artificial ellipticities induced by the first pixelization
effect can be as large as $\langle e^2 \rangle^{1/2} \gtrsim 10^{-2}$ 
if a characteristic size of objects (e.g., the FWHM) is smaller than 
twice of the pixel size. 
Whereas for objects with the characteristic size being
larger than three times of the pixel size, the RMS becomes negligibly
small ($\langle e^2 \rangle^{1/2} \lesssim 10^{-5}$). 
\item The second pixelization effect preferentially induces the $e_1$
 component (parallel to the grids). The reason of this is that 
pixels are square shaped and so the pixelization effect is,
in general, not axially symmetric but has some special directions. 
\item The size (e.g., RMS of $e$) of the shape deformation caused by the
  second pixelization effect depends on the object size.
It also strongly depends on the interpolation scheme for resampling and
on the grid size of the output pixels. 
If we set an upper limit of the RMS ellipticies by 
$\langle e^2 \rangle^{1/2} < 5 \times 10^{-3}$ for objects with 
FWHM$>2.5 \times l_{\rm pixel}$  (corresponding to FWHM$>0.5$ arcsec for 
the case of Suprime-Cam), the interpolation schemes 
passing the above condition are (see Figure \ref{fig:thetag}) 
the 5th order polynomial,
Lanczos3, Lanczos4 and sinc kernel (as far as among ones considered in
this paper). 
Adopting the grid refinement makes a great improvement.
Actually, if one adopts twice finer grid for output pixels, even
the linear polynomial can pass the above condition.
\item Resampling of a pixelized image with rotation induces a periodic 
concentric-circle-shaped pattern of artificial ellipticities in object
shapes.
The scale of the pattern is related to the pixel size and the rotation 
angle, $\phi$,  by $L_{\rm pattern} = l_{\rm pixel} / \tan\phi$.
\end{itemize}

Before closing this paper, we would like to make a comment on an impact
of the second pixelization effect on the actual weak lensing analysis 
using Suprime-cam data presented in Miyazaki et al. (2007). 
Miyazaki et al. (2007) carried out resampling adopting 
the 3rd order polynomial interpolation scheme and 
combined typically 4 dithered images\footnote{Combining dithered 
images reduces the RMS ellipticities roughly as
$\propto N^{-1/2}$ for $N$ dithered images (Rhodes et al. 2007), 
because an object falls onto a different sub-pixel position
in different exposures as a consequence of 
dithered exposures which results in different ellipticities with
basically random orientations.
Combining those images can mitigate the pixelization effects.}, 
thus for the images with
FWHM$\gtrsim 0.6$ arcsec (the typical PSF size), the RMS of 
ellipticities induced by the second pixelization effect should be well
below $10^{-2}$.
Whereas, the RMSs measured from stellar images are about a few $\times
10^{-2}$, 
therefore we may safely conclude that the second pixelization effect is
suppressed sufficiently, and is not a major source of
the artificial ellipticities in object shapes.

\bigskip

We would like to thank Richard Massey for valuable comments
and Nick Kaiser for making the IMCAT software
available.  
We would like to thank the anonymous referee for valuable and
constructive comments on the earlier manuscript which improve the paper.
This research was supported in part by the Grants--in--Aid from
Monbu--Kagakusho and Japan Society of Promotion of Science (15340065 and
17740116).
Numerical computations presented in this paper were carried out on computer
system at CfCA (Center for Computational Astrophysics) and at ADAC
(Astronomical Data Analysis Center) of the National
Astronomical Observatory Japan.



\begin{thebibliography}{99}

\bibitem[BS(2001)]{BS2001}
Bartelmann M., Schneider P.\ 2001, Phys. Rep., 340, 291

\bibitem[FM(1994)]{FM94}
Fort, B., Mellier, Y.\ 1994, A\&AR, 5, 239

\bibitem[Hamana et al (2003)]{Hamana+03}
Hamana, T., Miyazaki, S., Shimasaku, K., Furusawa, H., 
Doi, M., Hamabe, M., Imi, K., Kimura, M., Komiyama, Y., 
Nakata, F., Okada, N., Okamura, S., Ouchi, M., Sekiguchi, M., 
Yagi, M., Yasuda, N.\ 2003, ApJ, 597, 98

\bibitem[STEP1 (2006)]{STEP2006}
Heymans, C. et al. 2006, MNRAS, 368, 1323

\bibitem[Kaiser et al.(1995)]{ksb95}
Kaiser, N., Squires, G., Broadhurst, T.\ 1995, ApJ, 449, 460

\bibitem[STEP2 (2007)]{STEP2007}
Massey, R. et al. 2007, MNRAS, 376, 13

\bibitem[Mellier(1999)]{Mellier99}
Mellier, Y.\ 1999, ARA\&A, 37, 127

\bibitem[Suprime-Cam(2002)]{Suprime-Cam02}
Miyazaki, S., Komiyama, Y., Sekiguchi, M., Okamura, S.,
Doi, M., Furusawa, H., Hamabe, M., Imi, K., Kimura, M., 
Nakata, F., Okada, N., Ouchi, M., Shimasaku, K., Yagi, M.,
Yasuda, Naoki.\ 2002, PASJ, 54, 833

\bibitem[Suprime-Cam(2007)]{s33-p1}
Miyazaki, S., Hamana, T., Ellis, R. S., Kashikawa, N., Massey, R. J., 
Refregier, A.\ 2007, ApJ, 669, 714

\bibitem[Refregier(2003)]{Refregier03}
Refregier, A.\ 2003, ARA\&A, 41, 645 

\bibitem[Rhodes et al(2007)]{Rhodes+07}
Rhodes, D. J., Massey, R., Albert, J., Collins, N., Ellis, R. S., 
Heymans, C., Gardner, J. P., Kneib, J-P., Koekemoer, A., Leauthaud, A., 
Mellier, Y., Refregier, A., Taylor, J. E., Van Waerbeke, L.\ 2007, ApJS, 
172, 203

\end{thebibliography}
\end{document}